\documentclass[a4paper,10pt]{article}
\usepackage{amsmath}
\usepackage{amsfonts}
\usepackage{amssymb}
\usepackage{graphicx}
\begin{document}

\title{  \textsc{Slow roll inflation and BB mode angular  power spectrum of CMB}}

\author{N. Malsawmtluangi $ \footnote{e-mail: tei.naulak@uohyd.ac.in}$
        and
        P. K. Suresh $\footnote{e-mail: pkssp@uohyd.ernet.in}$ \\[.2cm]
{\small \it School of Physics, University of Hyderabad. }\\
{\small \it P. O. Central University, Hyderabad 500046. India.}}

\date{\empty}

\maketitle

\begin{abstract}
The BB-mode angular correlation power spectrum of CMB is obtained by considering  the  primordial gravitational waves in the squeezed vacuum state for various inflationary models  and  results are compared with the joint analysis of  the BICEP2/Keck Array and Planck 353 GHz  data.  
 The present results may constrain several models of inflation.
\end{abstract}

\section{Introduction}
Cosmic inflation is the most widely known scenario proposed for resolving several problems associated with the standard model of cosmology \cite{bbg, lindebook}.  A number of  inflationary models have been proposed  over  several decades \cite{bbg,lindebook,linde3, guth1,linde1, guth2,linde2,jgbell}.  
The  recent observations  on the cosmic microwave background (CMB) anisotropy  data may constrain many of the inflationary models  \cite{ade, ade2013, jmart1,jmart2, jmart3}. It is believed that  inflation seeded the  formation of the large scale structures in the universe. Inflation also predicts a nearly scale invariant spectrum for the scalar and tensor perturbations which occurred in the early universe. The tensor perturbations of cosmological  origin are  known as primordial gravitational waves  (GWs).

It is believed that the primordial gravitational waves have left its imprint on the cosmic microwave background. Therefore the primordial GWs can be studied with the aid of  CMB anisotropy and polarization.   The CMB   is polarized  in the early universe  due to the Thomson scattering. The density (scalar) fluctuations generate the $E$-mode polarization of the CMB, while the gravitational waves generate both $E$-mode and $B$-mode polarizations \cite{dodelson,dodelson2}. The  primordial gravitational waves are a unique source of $B$-mode of CMB and  its detection will help in understanding the inflation as well as  the primordial gravitational waves itself.
 
The gravitational waves  were generated during the inflation period  due to the zero-point quantum oscillations \cite{lpg1}. An initial  vacuum state (no graviton) can evolve into a multi-particle quantum state known as  the squeezed vacuum state \cite{lpg3},  which is a well-known state in the context of  quantum optics \cite{bls, jrk, cmc}. Therefore the  primordial gravitational waves  are believed  to exist  in  the  squeezed vacuum state. \cite{lpg2,lpg4,lpg5}. The  primordial gravitational waves are placed in  the squeezed vacuum  state   and its  effect    on the $BB$ mode angular correlation power spectrum of CMB is   studied with WMAP data  \cite{bgh2}. Recently, it is shown that  the   $BB$ mode angular power spectrum gets enhanced at its lower multi poles by considering   the primordial gravitational waves in  thermal  state \cite{prl}. These studies show that the primordial gravitational waves may exhibit both the squeezing as well as thermal features  and hence it is worthwhile to examine their combined effects  on the $BB$ mode angular correlation power spectrum in light of  the recent  joint BICEP2/Keck Array and Planck data.
 
 The  aim of the present work is to   study effect of primordial gravitational waves   in  the squeezed vacuum state 
   on the $BB$ mode angular  correlation power spectrum of CMB with various slow roll inflationary models.  Thus the obtained  $BB$ mode angular correlation power  spectrum of CMB   for the  squeezed vacuum as well as the joint  effect of squeezing and thermal cases  are compared with  the  joint BICEP2/Keck Array and Planck data.

\section{Tensor power spectrum in squeezed state}
The perturbed metric for  a flat Friedmann-Lema\^{i}tre-Robertson-Walker  universe can be written as
\begin{equation}
ds^2 = R^2(\tau)[-d\tau^2 + (\delta_{ij}+h_{ij})dx^idx^j],
\end{equation}
where $\delta_{ij}$ is the flat space metric   and $h_{ij}$ is the  tensor perturbation, $|h_{ij}|\ll\delta_{ij}$,  $ \partial_ih^{ij} = 0,\delta^{ij}h_{ij}=0$ and $d\tau = \frac{dt }{R}$ is the conformal time.

In quantum theory, the field $h_{ij}(\textbf{x},\tau)$ can be written in the Fourier mode as
\begin{eqnarray}\label{fourm}
  h_{ij}(\textbf{x},\tau)=\frac{D}{(2\pi)^\frac{3}{2}}\int_{-\infty}^{+\infty}\frac{d^3\textbf{k}}{\sqrt{2k}}\sum_{p=1}^2  
\left[ h_k^{(p)}(\tau) b_k^{(p)}e^{i\textbf{k}.\textbf{x}}\varepsilon_{ij}^{(p)}(\textbf{k}) 
  + h_k^{(p) \ast}(\tau)b_k^{(p) \dagger}e^{-i\textbf{k}.\textbf{x}}\varepsilon_{ij}^{(p) \ast}(\textbf{k}) \right], 
\end{eqnarray}
where $D=\sqrt{16\pi}l_{pl}$ is the normalization constant, $l_{pl}=\sqrt{G}$ is the Planck length and   $\textbf{k}$ is  the wave vector. The wave number is $k=(\delta_{ij}k^ik^j)^\frac{1}{2}$ and is related to wavelength, $ \lambda = \frac{2\pi R}{k}$.
The two polarization states $\varepsilon_{ij}^{(p)}$, $p=1,2$  satisfy the conditions
$ \varepsilon_{ij}^{(p)}\delta^{ij}=0, ~\varepsilon_{ij}^{(p)}k^i=0, ~\varepsilon_{ij}^{(p)}\varepsilon^{(p') ij} = 2\delta_{pp'},
 ~\varepsilon_{ij}^{(p)}(\textbf{-k})=\varepsilon_{ij}^{(p)}(\textbf{k}).$
These linear polarizations are  respectively known as   plus $(+)$ polarization and cross $(\times)$ polarization.

The creation ( $b_k^{(p) \dagger}$ ) and annihilation ($b_k^{(p)}$) operators   satisfy the relationships,\, $ [b_k^{(p)},b_{k'}^{(p') \dagger}]$ =$\delta_{pp'}\delta^3(k-k') $ and  $
[b_k^{(p)},b_{k'}^{(p')}]$ =$[b_k^{(p) \dagger},b_{k'}^{(p') \dagger}]=0.$
The evolution of these operators are governed by the Heisenberg equations of motion,
\begin{equation}\label{Hsa}
\frac{d}{d\tau}b_{k}^{\dagger}(\tau) = -i[b_k^{\dagger}(\tau),H], \,\, 
\frac{d}{d\tau}b_{k}(\tau) = -i[b_k(\tau),H].
\end{equation}
The initial vacuum state $|0\rangle$ is defined as
\[b_k^{(p)} |0\rangle = 0.\]
Under the Bogoliubov transformation  the creation and annihilation operators become  
\begin{equation}
b_{k}^{\dagger}(\tau) = u_k^{\ast}(\tau)b_{k}^{\dagger}(0) + v_k^{\ast}(\tau)b_{k}(0),
\end{equation}
\begin{equation}
b_{k}(\tau) = u_{k}(\tau)b_{k}(0) + v_{k}(\tau)b_{k}^{\dagger}(0),
\end{equation}
where  $b_{k}^{\dagger}(0)$ and $b_{k}(0)$ are the initial values of the operators, $u_k(\tau)$ and $v_k(\tau)$ are complex functions and they  satisfy the condition
\[|u_k|^2-|v_k|^2=1.\] 
The coupling of the mode functions $h_k (\tau)$ with $R(\tau)$ gives 
\begin{equation}\label{mf}
h_k^{(p)}=\frac{\chi_k^{(p)}}{R},
\end{equation}
where $\chi_k^{(p)}$  can have the following form \footnote {Since we take the contribution from each polarization to be the same, here onwards we drop the superscript $(p)$. }
\begin{equation}\label{mff}
\chi_k(\tau) = u_k(\tau) + v_k^{\ast}(\tau),
\end{equation}
which satisfies the equation of  motion
\begin{equation}\label{eom}
\chi_k'' + \left(k^2-\frac{R''}{R}\right)\chi_k = 0,
\end{equation}
where prime  indicates the derivative with respect to the conformal time $\tau$.

Two-point correlation function of the  tensor perturbation is given by:
\begin{equation} \label{tpc}
\langle h_k h^{\ast}_{k'}\rangle = \frac{2\pi^2}{k^3}P_T(k)\delta^3(\textbf{k}-\textbf{k}'),
\end{equation}
where the angle bracket denotes ensemble average and  $P_T$ is known as the tensor  power spectrum. 

Using eqs. (\ref{fourm}) and (\ref{mf}), taking the contribution from each polarization to be the same, 
\begin{equation}\label{field}
h(\textbf{x},\tau) = \frac{D}{R(\tau)(2\pi)^\frac{3}{2}}\int_{-\infty}^{+\infty}d^3\textbf{k} [\chi_k(\tau) b_k + \chi_k^{\ast}(\tau)b_k^{\dagger}]e^{i\textbf{k}.\textbf{x}}.
\end{equation}

The primordial gravitational waves are created due to the   zero-point quantum oscillations in the early universe \cite{lpg1,lpg6}. The  initial vacuum state with no graviton evolves into multi-particle quantum state through parametric amplification.  Hence the primordial GWs are  possible to be considered in the squeezed vacuum state \cite{lpg2, nakamura}. 

The squeezed vacuum state is defined as \cite{bls,dfw}
\begin{equation}
|\zeta\rangle = Z(\zeta)|0\rangle,
\end{equation}
where $Z(\zeta)$ is the single mode squeezing operator and is given by
\begin{equation}
Z(\zeta) = \exp\left[\frac{1}{2} \zeta^{\ast} d^2 - \frac{1}{2} \zeta d^{\dagger 2}\right],
\end{equation}
where $\zeta =r_s e^{i \gamma}$ is a complex number, $r_s$ is the squeezing parameter and $\gamma$ is the squeezing angle. The unitary transformations of the squeezing operator $S$ on the annihilation and creation operators lead to:
\begin{eqnarray}\label{qso}
 \nonumber && Z^{\dagger}(\zeta) \, d \,Z(\zeta) = d \cosh r_s -d^{\dagger} e^{i\gamma} \sinh r_s,  \label{32}\\
 && Z^{\dagger}(\zeta) \, d^{\dagger} \, Z(\zeta) = d^{\dagger}\cosh r_s -d e^{-i\gamma} \sinh r_s. \label{33}
\end{eqnarray}

Using eqs. (\ref{fourm}) and (\ref{qso}), the two point correlation  for the tensor perturbation in the squeezed vacuum is obtained as
\begin{eqnarray} \label{tpg} 
   \langle h_kh_{k'}^{\ast} \rangle = \frac{D^2}{R^2}  \left[ (1+ 2\sinh^2 r_s)|\chi_k|^2 
  +\frac{1}{2} \sinh2r_s (\chi_k^2 e^{i\gamma} + \chi_k^{\ast 2}e^{-i\gamma}) \right]    \delta^3 (\textbf{k}-\textbf{k}'). 
\end{eqnarray}
 Thus from eqs. (\ref{tpc}) and  (\ref{tpg}), we get the tensor power spectrum in the squeezed vacuum state as
\begin{eqnarray} \label{sqzspc}
 P_T(k)= \frac{k^3}{2\pi^2} \frac{D^2}{R^2} \left[(1+ 2\sinh^2 r_s)|\chi_k|^2
+\frac{1}{2} \sinh2r_s (\chi_k^2 e^{i\gamma} + \chi_k^{\ast 2}e^{-i\gamma})\right].
\end{eqnarray}

In the case of quasi de Sitter universe during inflation, $R(\tau)=\frac{-1}{H\tau (1-\epsilon)}$, where  the slow roll parameter $\epsilon$ is related  to  the scalar field potential $V$  and is  given by, $\epsilon = \frac{m_{pl}^2}{2}\left(\frac{V'}{V}\right)^2$. For small $\epsilon$, $\vartheta = \frac{3}{2}+\epsilon$, and $n_T = -2\epsilon = 3-2\vartheta$.

If the slow roll parameter  $\epsilon$ is  considered  as constant, then eq. (\ref{eom}) gets modified as
\begin{equation} \label{mudd}
\chi''_k + \left[k^2 -\frac{1}{\tau^2}\left(\vartheta^2 - \frac{1}{4}\right)\right]\chi_k =0.
\end{equation}
The general solution for eq. (\ref{mudd}) is given by
\begin{equation} \label{gsol}
\chi_k(\tau) = \sqrt{-\tau}[C_1 (k)H^{(1)}_{\vartheta}(-k\tau)+C_2 (k)H^{(2)}_{\vartheta}(-k\tau)],
\end{equation}
where $H^{(1)}_{\vartheta}$ and $H^{(2)}_{\vartheta}$ are respectively the Hankel functions of the first and second kind, and $C_1$ and $C_2$ are the constants of integration. For 
within the horizon ($k>>RH$), the  approximate  solution  is given by 
\[\chi_k(\tau)=\frac{1}{\sqrt{2k}}e^{-ik\tau}.\]
Using the above approximate  solution, one can obtain
\begin{eqnarray}
C_1(k)=\frac{\sqrt{\pi}}{2}\exp\left[i\left(\vartheta+\frac{1}{2}\right)\left(\frac{\pi}{2}\right)\right], \, \, 
C_2(k)=0.
\end{eqnarray}
Therefore, for long wavelength limit ($k<<RH$), eq. (\ref{gsol}) leads to
\begin{equation} \label{musol}
\chi_k(\tau)=e^{i\left(\vartheta-\frac{1}{2}\right)\left(\frac{\pi}{2}\right)}2^{\vartheta-\frac{3}{2}}\frac{\Gamma(\vartheta)}{\Gamma{\left(\frac{3}{2})\right)}}\frac{1}{\sqrt{2k}}(-k\tau)^{\frac{1}{2}-\vartheta}.
\end{equation}

Using eq. (\ref{musol}) in eq. (\ref{sqzspc}), the tensor power spectrum in the superhorizon limit ($k<<RH$) is obtained as,
\begin{eqnarray}\label{gwsvs}
 P_T(k)= A_T(k_0) \left(\frac{k}{k_0}\right)^{n_T}  [1+2\sinh^2 r_s 
  + \sinh2r_s \cos \left(\gamma + (2-n_T)\frac{\pi}{2} \right) ],
\end{eqnarray}
where $A_T(k_0) = D^2 \left(\frac{H_{k0}}{2\pi}\right)^2$ is the normalization constant and  $H_{k0}$ is the Hubble parameter at $RH=k_0$ during the inflation, $k_0$ being the pivot frequency.
 
\section{Inflationary models and tensor power spectrum}
 In most models of inflation \cite{linde3},  a homogeneous scalar field, called inflaton,  is  considered as candidate  for the inflation.  The inflaton field  $\phi$ is governed by  the equation of motion given by
\begin{equation}
\ddot{\phi} + 3H \dot{\phi} + V' =0,
\end{equation}
where dot and prime indicate derivatives with respect to time ($t$) and the field ($\phi$) respectively. The Hubble parameter $H$ is determined by the energy density of the inflaton field,\[ \rho_{\phi} = \frac{\dot{\phi}^2}{2}+V,\] thus  the Friedmann equation can be written as
\begin{equation}\label{fdmeq}
H^2 = \frac{1}{3 m^2_{pl}}\left( \frac{1}{2} \dot{\phi}^2 + V(\phi)\right).
\end{equation}
In the slow-roll limit, the energy density of the inflaton field is dominated by its potential, $\frac{\dot{\phi}^2}{2} \ll V$.  From eq. (\ref{fdmeq}), the Hubble parameter and the inflaton potential are related by
\begin{equation}
H^2 \simeq \frac{V}{3m^2_{pl}}.
\end{equation}
 This condition is characterized by the slow-roll parameters which are defined in terms of the inflaton potential $V$ and its derivatives as \cite{guth1}
\begin{eqnarray}
\epsilon &\equiv & \frac{m_{pl}^2}{2}\left(\frac{V'}{V}\right)^2, \nonumber \\ \eta &\equiv & m^2_{pl}\left(\frac{V''}{V}\right),\\ \xi^2  &\equiv & m^4_{pl} \frac{V' V^{'''}}{V^2}, \nonumber \\ \sigma^3 &\equiv & m^6_{pl} \frac{V'' V^{''''}}{V^3}, \nonumber
\end{eqnarray}
and so on. Inflation lasts as long as the slow-roll conditions are satisfied, i.e., $\epsilon \ll 1$ and $|\eta|\ll 1$. The duration of inflation is characterized by the e-fold number, $N$, which can be written in terms of the inflaton potential,
\begin{equation}\label{efold}
N \simeq \frac{1}{m^2_{pl}} \int^{\phi}_{\phi_{end}} \frac{V}{V'} d\phi.
\end{equation}
In the slow-roll approximation, the power spectrum  of the scalar perturbations ($P_S$) and the tensor perturbations ($P_T$) generated outside the horizon are respectively  given in terms of the potential  \cite{skuro4, skuro1}
\begin{eqnarray}
P_S &\simeq & \frac{1}{12 \pi^2 m^6_{pl}}\frac{V^3}{V'^2} |_{k=RH},\\
P_T &\simeq & \frac{1}{3 \pi^2 m^4_{pl}}V |_{k=RH},
\end{eqnarray}
where $k=RH$ indicates that $H$ and hence, $V$ is evaluated at the time when the mode with wave number $k$ crosses the horizon. For all calculations, we take the scalar power spectrum to be $P_S= 2.43 \times 10^{-9}$.

The tensor-to-scalar ratio can be written in terms of the parameter $\epsilon$ as
\begin{equation}\label{r}
r \equiv \frac{P_T (k)}{P_S (k)} \simeq 16\epsilon.
\end{equation}
This parameter is often used to characterize the amplitude of the tensor perturbation at the CMB scale. Physically, {\it r} is a measure of the slope (of the quantum hill) down which the scalar field is rolling. Since inflation predicts a nearly scale invariant spectrum, the slope is small but not flat. Hence, {\it r} is small and can differentiate between the many inflation models.

Next, we obtain  the slow roll parameter and tensor spectral index  corresponding  to  various  slow roll inflation models given below.  For this purpose, 
we use  the  e-fold number  $N = 60$, for all the inflationary models under the present work \cite{wzhao}.

\subsection{Quadratic Chaotic Inflation}
The chaotic inflation model assumes that the scalar field rolls down  its potential and rests at its vacuum state for a while,  then after getting displaced due to some fluctuations, rolls back to its true vacuum state and the same mechanism repeats itself.
For quadratic chaotic inflation, the scalar field has the potential given by \cite{guth1}
\begin{equation}\label{ch2pot}
V(\phi) = \frac{1}{2}m^2 \phi^2,
\end{equation}
where $m$ is the mass of the  inflaton field and is taken to be $m= 1.53 \times 10^{13}$ GeV.

 Using eqs. (\ref{efold}) and (\ref{ch2pot}), 
 the slow roll parameter and index of the tensor power spectrum    for the quadratic chaotic inflationary model are obtained  as:
\begin{eqnarray}
 \epsilon_H &=& 8.26 \times 10^{-3},\nonumber \\ 
 n_{T_H} &\simeq & -1.65 \times 10^{-2}.
\end{eqnarray} 
The corresponding 
 tensor-to-scalar ratio is obtained  as, $r=0.132$.

\subsection{Quartic Chaotic Inflation}
The quartic chaotic inflation model suggests the existence of the curvaton which generates the curvature fluctuations during inflation after the inflaton field has decayed, while it does not drive the inflation itself. The model has the potential given by \cite{linde1}
\begin{equation}
V(\phi)=\frac{1}{4} \lambda \phi^4,
\end{equation}
where normalization gives the self-coupling of the scalar field ($\phi$) as $\lambda =5.94 \times 10^{-14}$ GeV, and hence 
\begin{eqnarray}
\epsilon_H &=& 1.626 \times 10^{-2},\nonumber \\ 
n_{T_H} &\simeq & -3.25 \times 10^{-2}. 
\end{eqnarray}
The 
tensor-to-scalar ratio is  for this model is found as $r=0.26$.

\subsection{New Inflation}
The new inflation model \cite{guth2} is based on the Coleman-Weinberg potential\cite{coleman, turner},
\begin{equation}
V(\phi) = \frac{1}{4} \lambda \phi^4 \left(\ln \frac{\phi}{\sigma}-\frac{1}{4}\right) + \frac{\lambda \sigma^4}{16}.
\end{equation}
In this model, the parameters are taken to be $\sigma=10m_{pl}$, where $m_{pl}$ is Planck's mass, and $\lambda = 2.36 \times 10^{-14}$ GeV. $\sigma$ is the vacuum expectation value of the scalar field at minimum and $\lambda$ is the quadratic self coupling of the field. 
Therefore we get
\begin{eqnarray}
\epsilon_H &=& 1.16 \times 10^{-2},\nonumber \\
 n_{T_H} &\simeq & -2.32 \times 10^{-2}. 
\end{eqnarray}
The 
tensor-to-scalar ratio in this case is obtained as $r=0.186$.

\subsection{Hybrid Inflation}
In this model, the inflation is driven by two scalar fields, in which one of the fields ($\phi$) is responsible for the normal slow-roll inflation while the other field ($\sigma$) triggers the end of inflation. The potential for this model is \cite{linde2, jgbell}
\begin{equation}
V(\phi)=\frac{1}{4\lambda}(M^2 - \lambda \sigma^2)^2+ \frac{1}{2}m^2\phi^2 + \frac{1}{2}g^2\phi^2\sigma^2,
\end{equation}
where the parameters are taken to be as follows:  $g=8 \times 10^{-4}$, $\lambda=1$ are the self-coupling constants of the inflaton field and the trigger field respectively, and the masses of the fields are $m=1.5 \times 10^{-7} m_{pl}$, $M =1.21 \times 10^{16}$ GeV respectively.

The $\phi$ field drives the inflation, thus determining the period of inflation while the $\sigma$ field determines the rate of inflation.
Therefore we get:
\begin{eqnarray}
\epsilon_H &=& 2.65 \times 10^{-4}, \nonumber \\ n_{T_H} &\simeq & -5.3 \times 10^{-4}.
\end{eqnarray}
   The tensor-to-scalar ratio in this case is found as $r=4.24 \times 10^{-3}$.

\section{BB-mode angular power spectrum of  CMB }

The CMB radiation can be polarized due to the   gravitational waves   \cite{hwayne} called the $B$-mode \cite{kosowsky, ytlin} of CMB.  

The angular correlation power spectrum of the $B$-mode CMB is given by \cite{usel,baskaran}
\begin{eqnarray}
  \frac{C_l^{BB}}{(4\pi)^2} = \int dk  \, k^2 \, P_T(k) 
  \left| \int_0^{\tau_H} d\tau g(\tau) h_k(\tau) \left[2j'_l(x) +\frac{4j_l(x)}{x}\right]\right|^2,
\end{eqnarray}
where $g(\tau)=\kappa e^{-\kappa}$ is the visibility function, $\kappa$ is the differential optical depth for the Thomson scattering, $x=k(\tau_H -\tau)$ and $j_l$ is the spherical Bessel function.

The 
$BB$ mode correlation angular spectrum  of CMB is obtained  for the  gravitational waves in the squeezed vacuum and thermal vacuum states for  
the   quadratic chaotic, quartic chaotic, new inflation and hybrid 
 inflationary models.  
The angular power spectra for the different inflation models are generated using the CAMB code with the tensor spectral index $n_T$ corresponding  to each inflationary model. For all the cases, the optical depth is taken as $\kappa = 0.08$, the pivot wave number for tensor modes is taken as $k_0 =0.002$ Mpc$^{-1}$ and that for scalar mode it is  $k_0 =0.05$ Mpc$^{-1} $. The  tensor to scalar ratio used for  each inflationary model is taken  from  the previous section where their values are computed.  The obtained  $BB$ mode angular power spectrum  for various values of  the squeezing parameter and temperature  are compared with the limit of  the BICEP2/Keck Array and Planck 353 GHz joint analysis data. The implemented  limit  (BK x BK - $\alpha$ BK x P)/(1 - $\alpha$) at $\alpha = \alpha_{fid} = 0.04$ is evaluated   \cite{bkp} from the auto-spectra and cross-spectra of the combined BICEP2/Keck 150 GHz maps and Planck 353 GHz maps to clean out the dust contribution, BK x BK indicates the BICEP2/Keck auto-spectra at 150 GHz and BK x P indicates the cross-spectra of BICEP2/Keck maps at 150 GHz and Planck maps at 353 GHz. This combination  is taken after the subtraction of the dust contribution which is 0.04 times as much in the BICEP2 band as it is in the Planck 353 GHz band (see ref.\cite{bkp} for details).

\begin{figure}[]
 \includegraphics[scale=0.4]{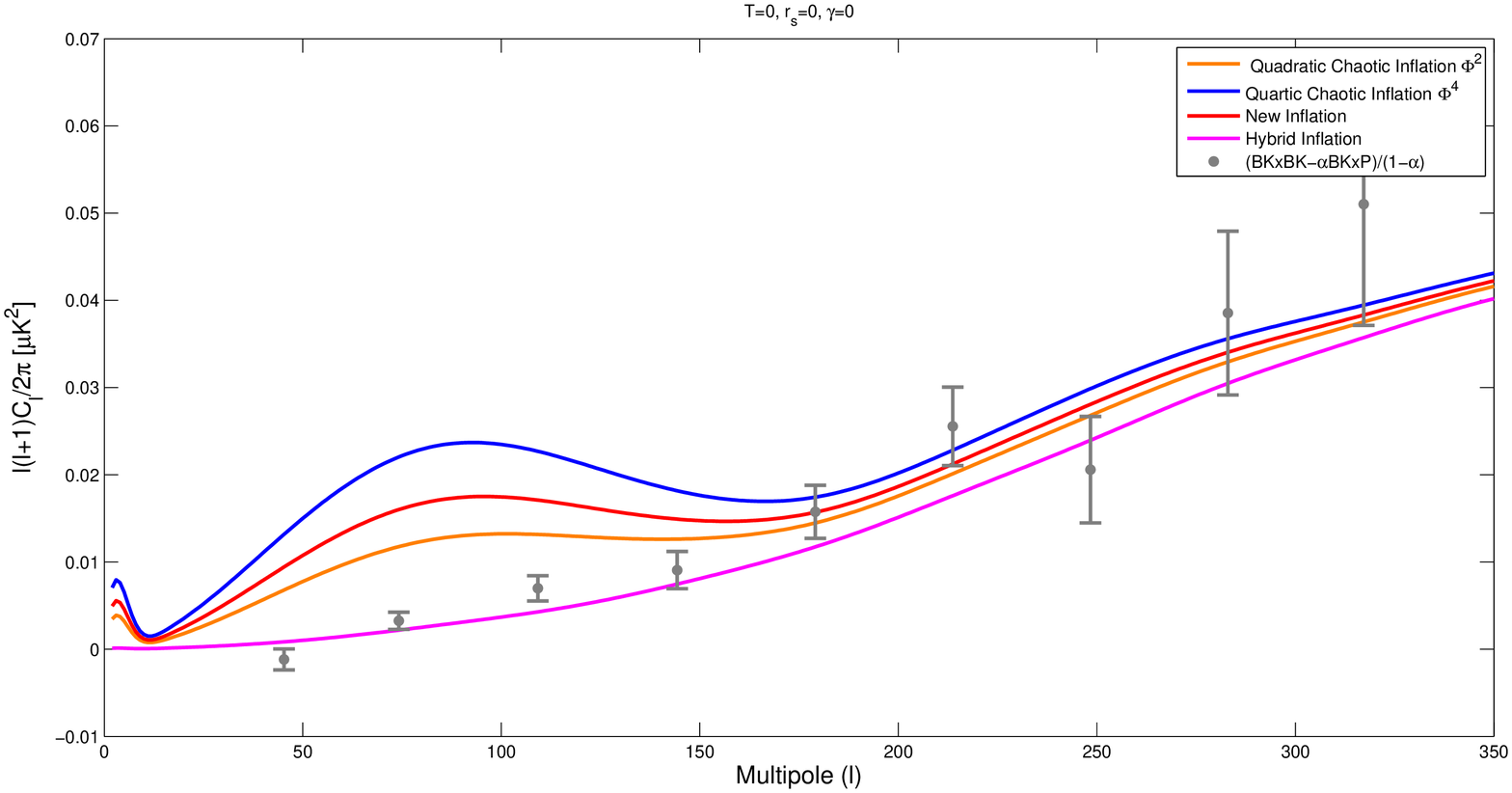}
\caption{ $BB$-mode angular spectra for various  inflation models with  lensing  for  zero squeezing  effect   with BICEP2/Keck and Planck joint analysis result \label{f01}}
 \end{figure}

\begin{figure}[]
  \includegraphics[scale=0.4]{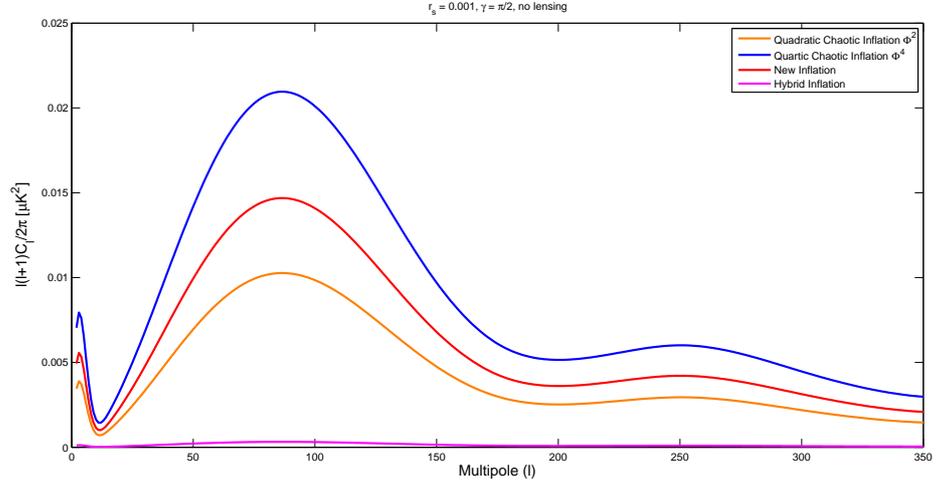}
\caption{  $BB$-mode angular spectra for  various inflation models with  unlensing for squeezing parameter $r_s = 0.001$  and squeezing angle $\gamma = \pi/2$  \label{f02}}
\end{figure}

\begin{figure}[]
  \includegraphics[scale=0.4]{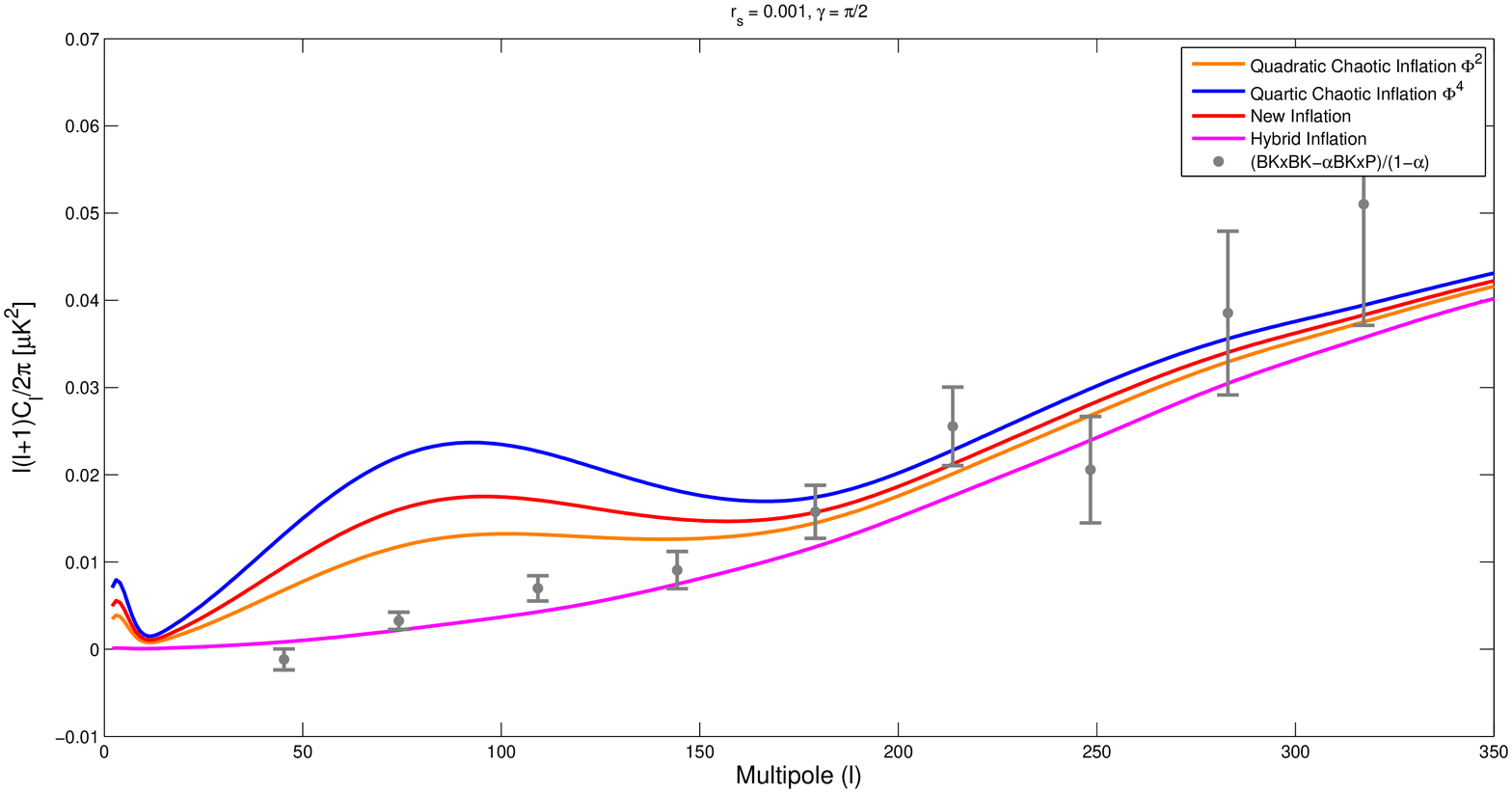}
\caption{  $BB$-mode angular spectra for  various inflation models with  lensing for squeezing parameter $r_s = 0.001$  and squeezing angle $\gamma = \pi/2$  with  BICEP2/Keck and Planck joint analysis result \label{f03}}
\end{figure}

\begin{figure}[]
  \includegraphics[scale=0.4]{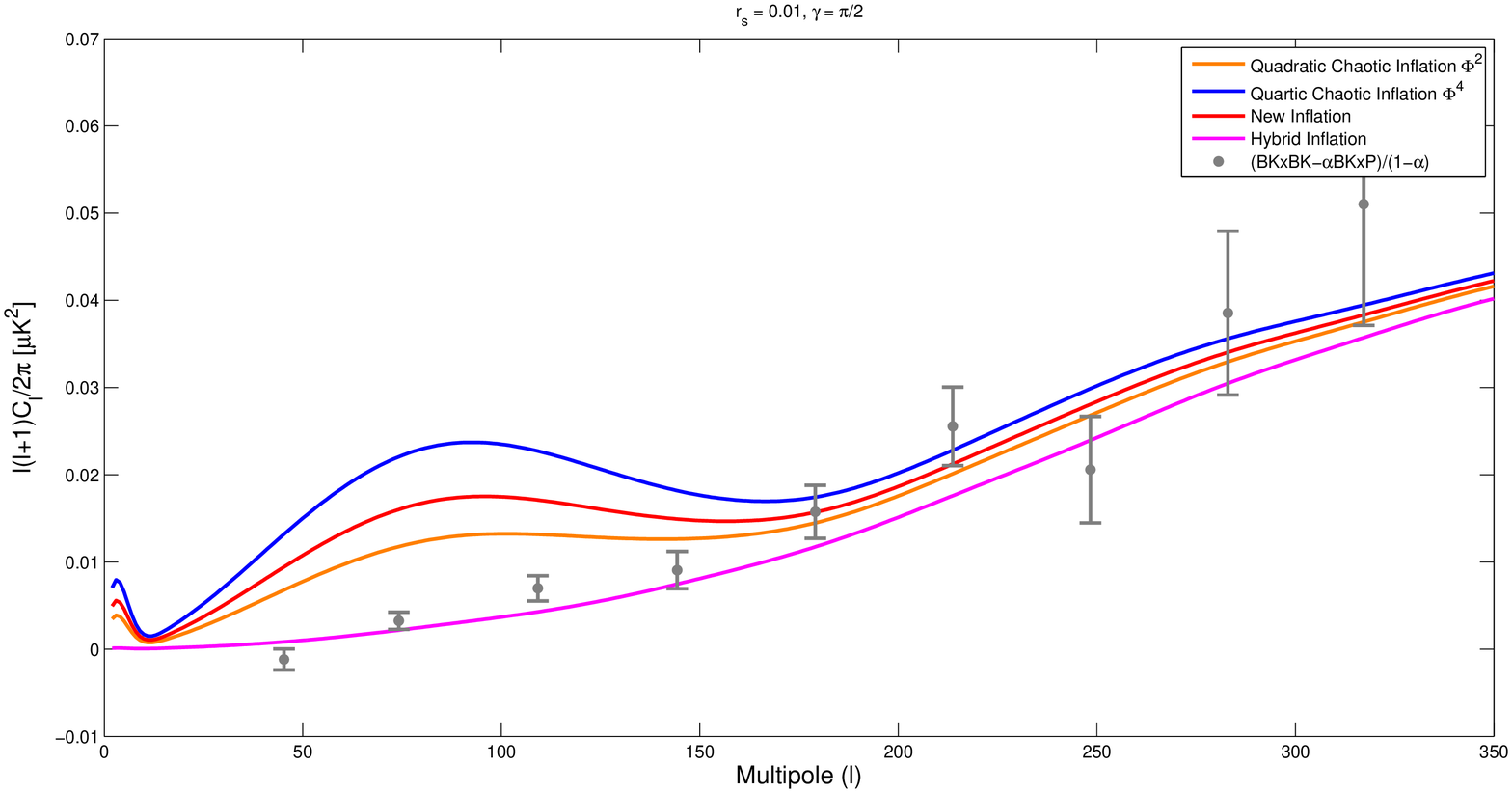}
\caption{  $BB$-mode angular spectra for various inflation models with lensing for  squeezing parameter $r_s = 0.01$ and squeezing angle $\gamma = \pi/2$  with  BICEP2/Keck and Planck joint analysis result \label{f04}}
\end{figure}

\begin{figure}[]
  \includegraphics[scale=0.4]{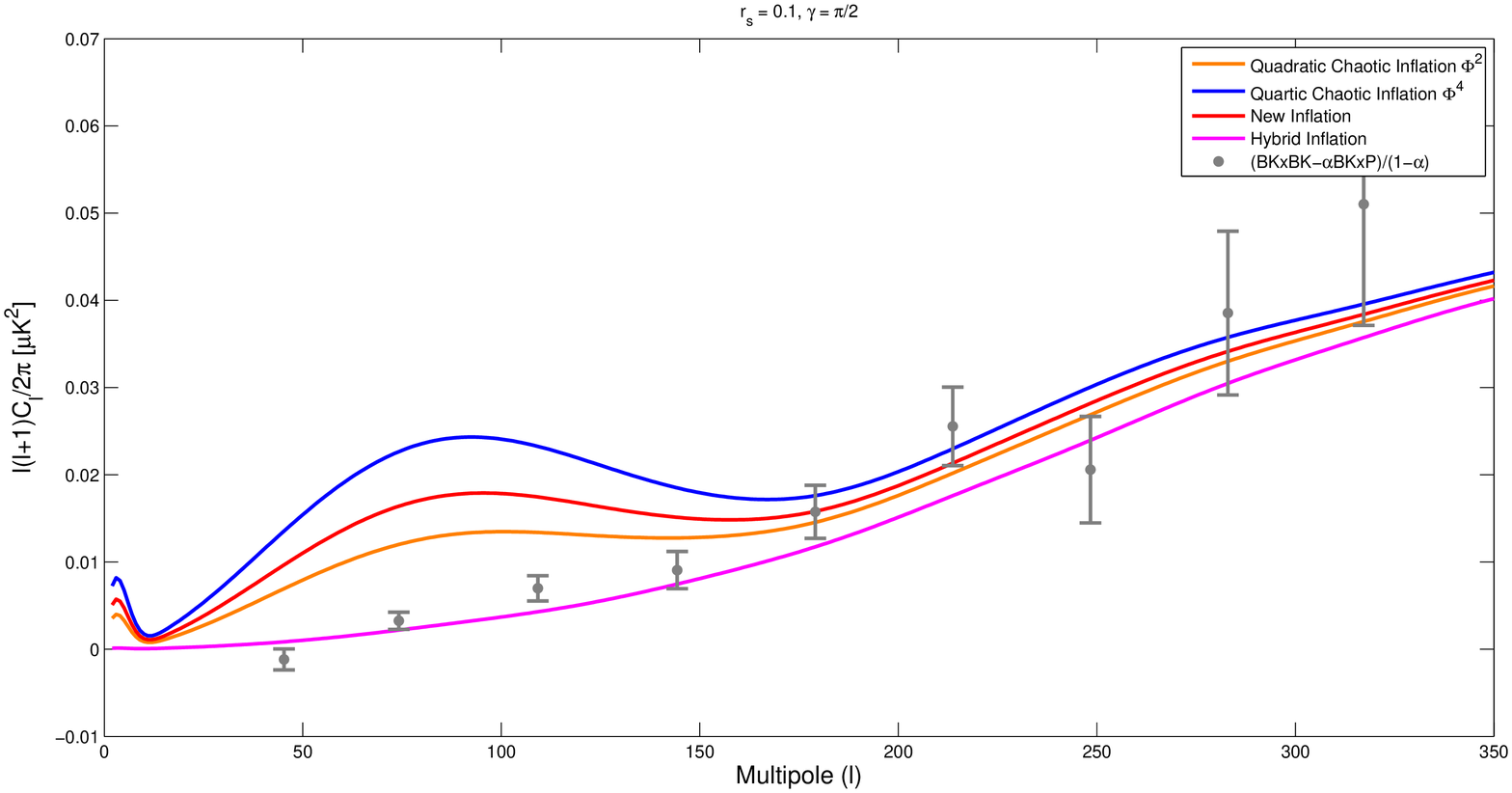}
\caption{ $BB$-mode angular spectra for various inflation models with lensing for squeezing parameter $r_s = 0.1$ and squeezing angle $\gamma = \pi/2$   BICEP2/Keck and Planck joint analysis result \label{f05}}
\end{figure}

\begin{figure}[]
  \includegraphics[scale=0.4]{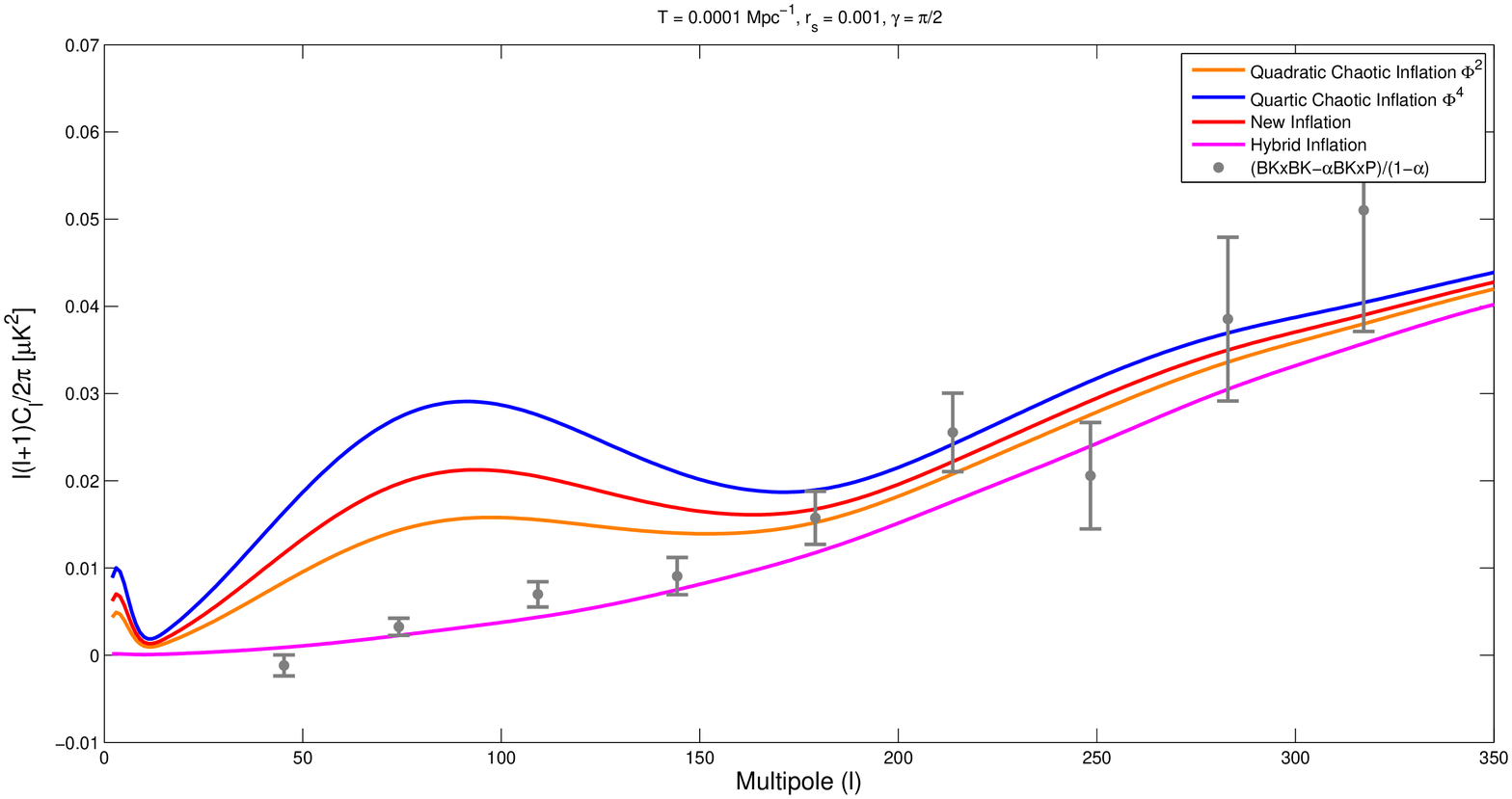}
\caption{  $BB$-mode angular spectra for various inflation models with lensing for  temperature  $T = 0.0001$ Mpc$^{-1}$ and squeezing parameter $r_s = 0.001$ 
and squeezing angle $\gamma = \pi/2$ with  BICEP2/Keck and Planck joint analysis result \label{f06}}
\end{figure}

\begin{figure}[]
  \includegraphics[scale=0.4]{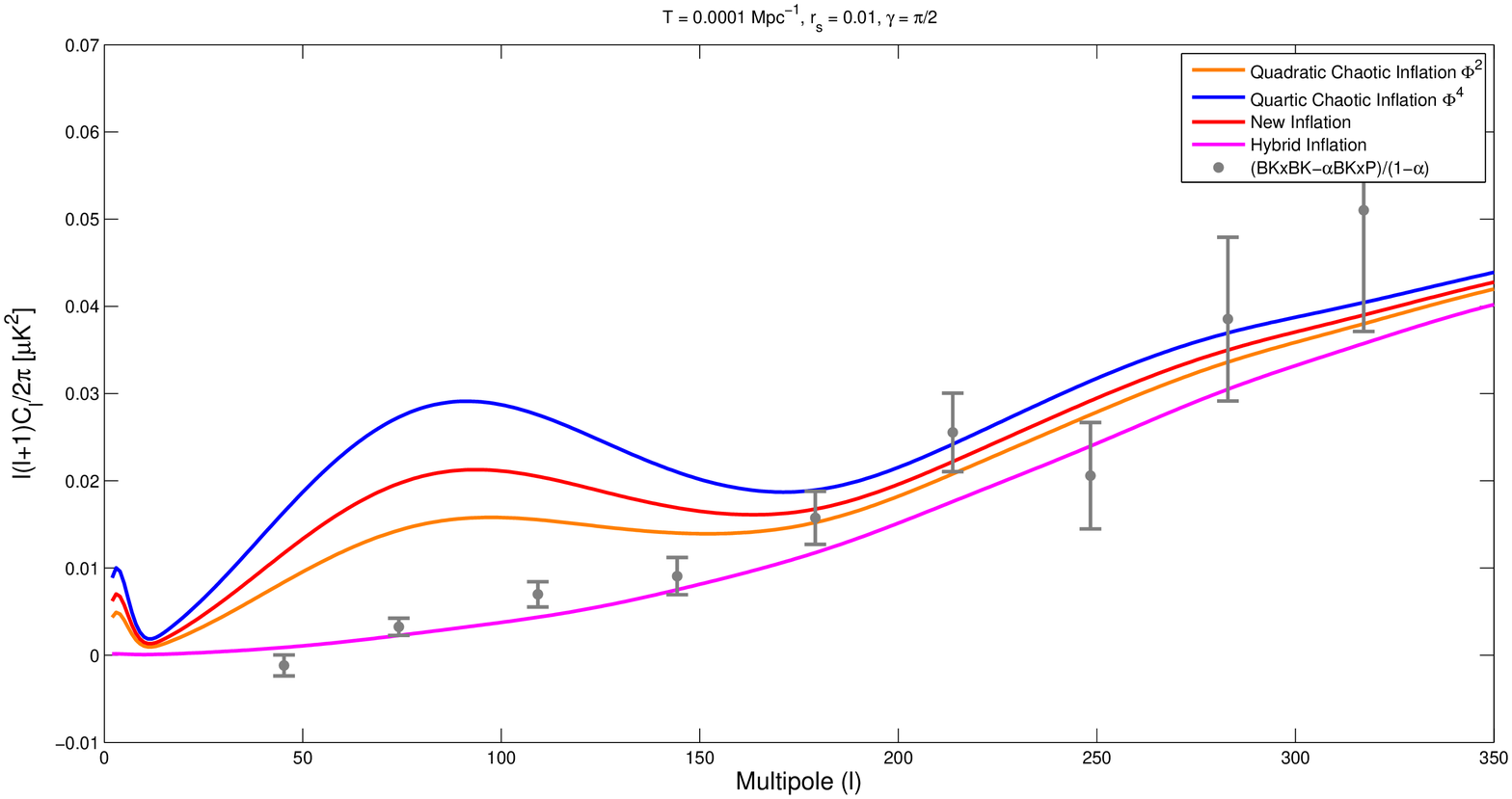}
\caption{  $BB$-mode angular spectra for various inflation models with lensing for temperature  $T = 0.0001$ Mpc$^{-1}$ and  squeezing parameter $r_s = 0.01$  and squeezing angle $\gamma = \pi/2$with BICEP2/Keck and Planck joint analysis result \label{f07}}
\end{figure}

\begin{figure}[]
  \includegraphics[scale=0.4]{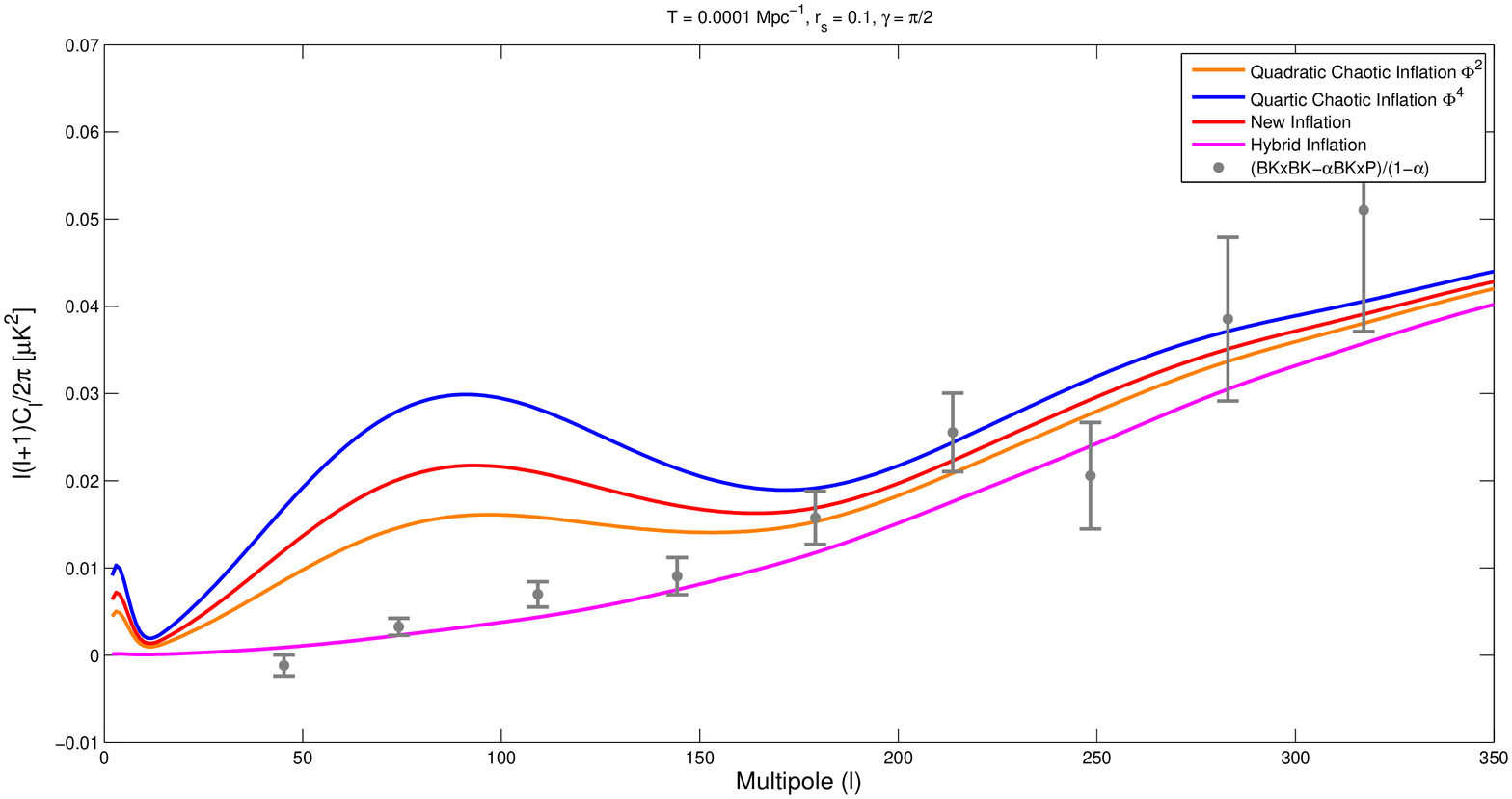}
\caption{  $BB$-mode angular spectra for various inflation models with lensing for temperature  $T = 0.0001$ Mpc$^{-1}$ and squeezing parameter $r_s = 0.1$ 
and squeezing angle $\gamma = \pi/2$  with  BICEP2/Keck and Planck joint analysis result \label{f08}}
\end{figure}

\begin{figure}[]
  \includegraphics[scale=0.4]{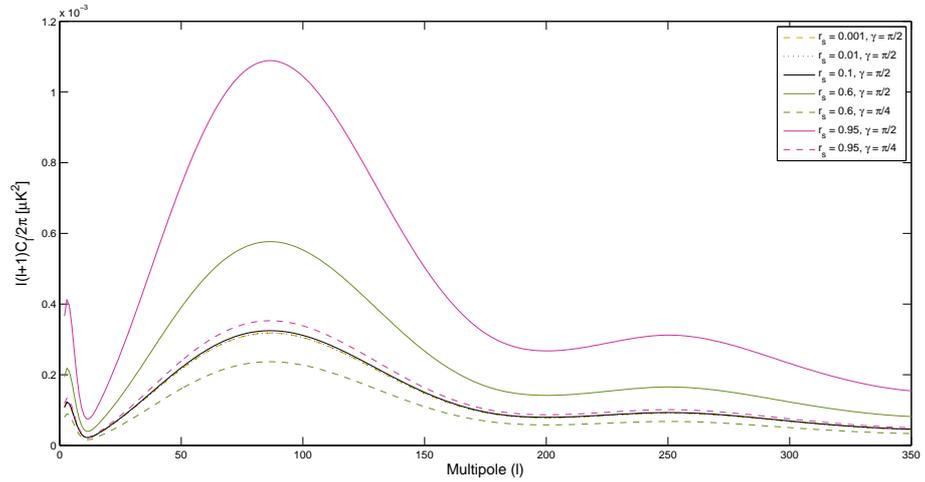}
\caption{  $BB$-mode angular spectrum   for  the hybrid  inflation model  with unlensing for various values of  squeezing parameter  and angles  \label{f09}}
\end{figure}

\begin{figure}[]
  \includegraphics[scale=0.4]{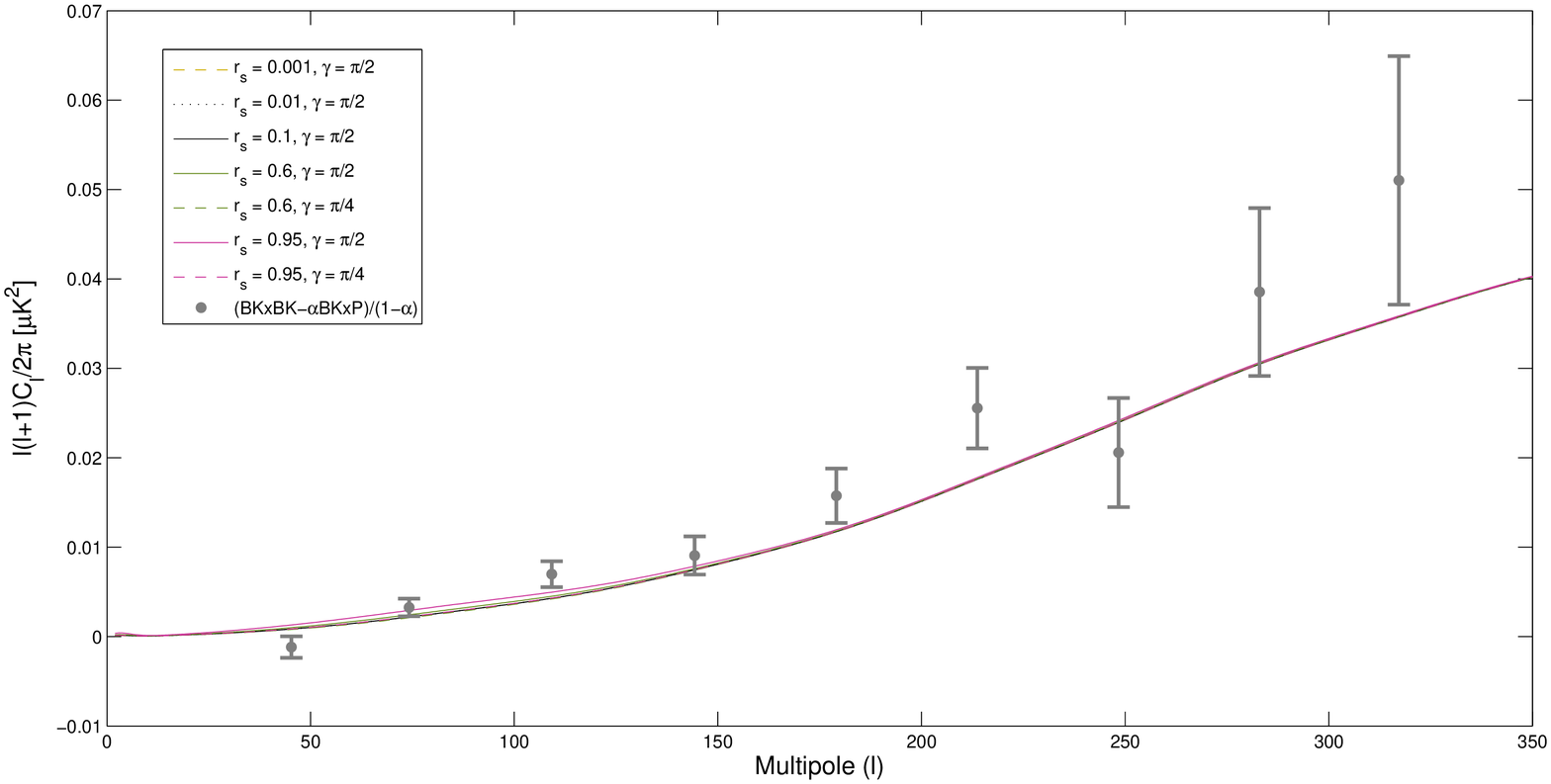}
\caption{ $BB$-mode angular spectrum for the hybrid  inflation model with lensing for various values of  squeezing parameter  and angles with BICEP2/Keck and Planck joint analysis result \label{f10}}
\end{figure}

The
 $BB$-mode correlation angular spectra   for the different inflation models for   zero squeezing  with lensing effect are  obtained and their corresponding plots are  given in fig.(\ref{f01}). The $BB$ mode angular spectrum   for  various inflationary models for
  the squeezing, thermal  and their combined effects  with lensed and  unlensed cases are also obtained and the results  plots are  in  figs. (\ref{f02}-\ref{f10}).   
It can be seen that
  for various values of the  squeezing parameter and squeezing angle,  the quadratic chaotic, quartic chaotic and new inflation models are out  of the limit  of BICEP2/Keck Array and Planck 353 GHz joint  data.    For the thermal effect with various values of the squeezing parameter and squeezing angle, the quadratic chaotic, quartic chaotic and new inflation models are also not found  within the limit of BICEP2/Keck Array and Planck 353 GHz joint  data. 
  
   It can be observed  that the  $BB$ mode angular correlation power spectrum corresponding to the hybrid inflation model for various values of squeezing parameter and squeezing angle is found within the limit of  the  BICEP2/Keck and Planck joint analysis data at $l \simeq 145$, even though the squeezing and  thermal effects enhances the angular correlation  power spectrum moderately. Thus the analysis of the present results  
  show that the hybrid inflation model is most favorable by  considering the primordial gravitational waves  in presence  of  squeezing thermal and effects.

\section{Conclusion}
The primordial  gravitational waves  are very important in cosmology.  It is believed that the origin of the primordial gravitational  waves are  due to inflation mechanism.   Thus the primordial gravitational waves are placed in a special quantum state called the squeezed vacuum state, a well known state in  quantum optics. The gravitational waves have not been detected directly yet but its effect is expected to be  observed  through  the  $B$ mode of CMB.
If this is the case, then the role of squeezing effect is also  expected to  reflect  on the $BB$ mode angular correlation power spectrum of CMB.

The $BB$-mode correlation  angular power spectrum for various slow-roll inflationary models are studied by considering the primordial gravitational waves  in the squeezed vacuum state.  The obtained angular spectrum  for the squeezing, thermal  as well as their combined effects  are compared with the  BICEP2/Keck Array and Planck 353 GHz joint analysis data.  
 The comparative  study of the  $BB$ mode angular spectrum obtained for the various inflationary models   with the BICEP2/Keck Array and Planck 353 GHz joint analysis data show that the hybrid inflation model is most favorable  with the primordial GWs in the squeezed vacuum state.  The $BB$  mode angular correlation power spectrum can be  obtained even with higher or lower values of the squeezing parameters and  thermal parameter but the results do not alter the present conclusions.
  Note that the $BB$ mode angular correlation power spectrum for the vacuum case can be recovered  in  absence of     squeezing and  thermal effects.

Further, the result of the joint analysis data does not rule out the existence of gravitational waves in the squeezed vacuum  or  thermal squeezed vacuum states which are in agreement with  previous studies. 

\section*{Acknowledgements}
N M  acknowledges the financial support of  Council of Scientific and Industrial Research (CSIR), New Delhi. P K S acknowledges financial  the support of   DST-SERB, New Delhi. We would like to thank Planck  and BICEP websites for the  data.\\

\section*{Appendix  A}

\section*{Thermal effect on GWs spectrum}\label{app}

After the Big Bang, the universe was  in the form of plasma of hot and dense matter. During this stage, the light particles like free electrons, due to Thomson scattering, acted as scattering centers for the surrounding radiation, keeping the universe at that time thermalized and opaque to radiations. 
 Among these particles and radiations which got thermalized and escaped after the recombination are the decoupled gravitons which left an imprint on the CMB anisotropy. Due to these thermalized gravitons, the gravitational waves are believed to be amplified by stimulated emission into the existing thermal background of gravitational waves which changes the spectrum of the waves by temperature-dependent factor \cite{prl}.

Assuming a vacuum state initially, the Fourier coefficients in Eq. (\ref{fourm}) satisfy the relations:
\begin{eqnarray}
\langle b_k^{\dagger}b_{k'}\rangle &=& \delta^3 (\textbf{k}-\textbf{k}'), \label{ckk} \\
\langle b_k b_{k'} \rangle &=& \langle b_k^{\dagger}b_{k'}^{\dagger} \rangle =0.
\end{eqnarray}

Time-dependent thermal state is defined as \cite{skoh,tfd}
\begin{equation}
|0 (\beta),\tau \rangle = \tilde{\emph{T}}(\theta_k) |0,\tau \rangle,
\end{equation}
where $\tilde{\emph{T}}(\theta_k) = exp [-\theta_k(\beta) \{\tilde{b}_k(\tau) b_k (\tau) -b_k^{\dagger}(\tau) \tilde{b}_k^{\dagger}(\tau)\}]$.

The temperature-dependent parameter $\theta (\beta)$ is defined by
\[\cosh \theta_k (\beta) = (1 - e^{-\beta k})^{-\frac{1}{2}},
\sinh \theta_k (\beta) = e^{-\frac{\beta k}{2}} (1 - e^{-\beta k})^{-\frac{1}{2}},\]
where $\beta = \frac{1}{T}$, $T$ is the temperature.

Then the time and temperature-dependent annihilation and creation operators through the Bogoliubov transformation become
\begin{eqnarray}
\nonumber b_k(\beta, \tau) &=& \tilde{\emph{T}} (\theta) b_k \tilde{\emph{T}}^{\dagger} (\theta) \\
&= &\cosh\theta_k (\beta)b_k(\tau)- \sinh\theta_k (\beta)\tilde{b}_{-k}^{\dagger} (\tau),\\
\nonumber \tilde{b}_k(\beta, \tau) &=& \tilde{\emph{T}} (\theta) \tilde{b}_k \tilde{\emph{T}}^{\dagger} (\theta)\\
&= &\cosh\theta_k (\beta)\tilde{b}_k(\tau)- \sinh\theta_k (\beta)b_{-k}^{\dagger} (\tau).
\end{eqnarray}
The Hermitian conjugate of the above equations give rise to similar equations for $b_k^{\dagger}(\beta, \tau)$ and $\tilde{b}_k^{\dagger} (\beta, \tau)$.

Thus, eq. (\ref{ckk}) modifies to \cite{wzhao}:
\begin{equation}
\langle b_k^{\dagger}b_{k'}\rangle = \left(1 + \frac{2}{e^{\frac{k_B}{T}}-1} \right) \delta^3 (\textbf{k}-\textbf{k}').
\end{equation}

The power spectrum for gravitational waves in the presence of thermal effect can then be written as

\begin{equation} \label{tvs}
P_T(k)=A_T(k_0) \left(\frac{k}{k_0}\right)^{n_T} \coth \left[\frac{k}{2T}\right].
\end{equation}
This can be used to compute the $BB$ mode correlation angular power spectrum of CMB  in the  thermal state.

\end{document}